\documentclass[sigconf,nonacm]{acmart}
\usepackage{amsmath,amsfonts}
\usepackage{stfloats}
\usepackage{float}
\usepackage{algorithmic}
\usepackage{graphicx}
\usepackage{textcomp}
\usepackage{xcolor}
\usepackage{url}
\usepackage{geometry}
\usepackage{subcaption}
\setcopyright{none}
\copyrightyear{}
\acmYear{}
\acmDOI{}
\acmISBN{}
\def\lowerboundcell{20k }
\def\upperboundcell{1.5M }
\AtBeginDocument{%
  }

\ifdefined\draft
  \newcommand{\hightideurl}{\texorpdfstring{Repository URL omitted for anonymous review.}{Repository URL omitted for anonymous review.}}
\else
  \newcommand{\hightideurl}{\texorpdfstring{\url{https://vlsida.github.io/HighTide} \url{https://github.com/VLSIDA/HighTide/}}{vlsida.github.io/HighTide; github.com/VLSIDA/HighTide}}
\fi
\ifdefined\draft
\author{Anonymous for Blind Review}
\else
 \author{Benjamin Goldblatt}
 \affiliation{%
   \institution{\textit{University of California, Santa Cruz}}
   \department{Computer Science \& Engineering}
   \city{Santa Cruz}
   \state{California}
   \country{USA}
 }
 \email{bgoldbla@ucsc.edu}
 \author{Paolo Pedroso}
 \affiliation{%
   \institution{\textit{University of California, Santa Cruz}}
   \department{Computer Science \& Engineering}
   \city{Santa Cruz}
   \state{California}
   \country{USA}
 }
 \email{ppedroso@ucsc.edu}
 \author{Farhad Modaresi}
 \affiliation{%
   \institution{\textit{University of California, Santa Cruz}}
   \department{Computer Science \& Engineering}
   \city{Santa Cruz}
   \state{California}
   \country{USA}
 }
 \email{semodar@ucsc.edu}
 \author{Ethan Sifferman}
 \affiliation{%
   \institution{\textit{University of California, Santa Cruz}}
   \department{Computer Science \& Engineering}
   \city{Santa Cruz}
   \state{California}
   \country{USA}
 }
 \email{esifferma@ucsc.edu}
 \author{Matthew R. Guthaus}
 \affiliation{%
   \institution{\textit{University of California, Santa Cruz}}
   \department{Computer Science \& Engineering}
   \city{Santa Cruz}
   \state{California}
   \country{USA}
 }
 \email{mrg@ucsc.edu}
\fi
 
\begin{document}
\title{HighTide: An Agent-Curated Open-Source VLSI Benchmark Suite}


\begin{abstract}
 We introduce HighTide\footnote{\hightideurl}, an evolving AI-assisted benchmark suite. Specifically, the contributions are: (i) a diverse open-source suite spanning multiple design languages and technology nodes, (ii) Bazel-based incremental RTL-to-GDS compilation with remote caching, (iii) AI-assisted design curation through twelve agent skills covering the design lifecycle, flow optimization, tool reference, and meta-maintenance, backed by per-design decision logs that serve as long-term memory of tuning rationale across the suite, and (iv) an infrastructure with RTL compilation verification for stable releases. The suite is publicly available and designed to grow with the open-source hardware ecosystem.
\end{abstract}


\maketitle
\setcounter{page}{1}


\section{Introduction}
\label{sec:intro}
Heterogeneous architectures have become dominant in system-on-a-chip (SoC) designs by integrating specialized processing units and accelerators. This is exemplified by mobile SoCs, which integrate central processing unit (CPU) cores, graphics processing units (GPUs), ML accelerators, video encoding/decoding engines, and cryptography accelerators~\cite{applem1}. These trends have seen increased adoption in the open-source hardware community, where designs representative of heterogeneous SoC components are becoming increasingly prevalent.

The hardware design language landscape is also evolving: SystemVerilog, Chisel, and Python-based generators are becoming alternatives to traditional Verilog. These higher-level languages enable parameterization that would be very difficult with hand-written register-transfer level (RTL) code, making them well-suited for configurability of components in modern SoCs. Though these languages ultimately compile down to Verilog for physical implementation, supporting them ensures that the suite can incorporate the growing body of open-source hardware developed outside of traditional Verilog. 

Existing benchmark suites have not been updated to reflect modern designs' complexity and structure. Such staleness can compromise the credibility of evaluations, leaving ML models that are trained on them to overfit on potentially outdated data and EDA tools to be benchmarked against designs that no longer reflect their current complexity. Models trained on a narrow design set will learn patterns specific to those designs, rather than generalizable circuit behaviors. 

In response to the above, we propose HighTide, a benchmark design suite that spans high-level languages through RTL-to-GDS, with diverse designs and infrastructure for tracking upstream updates. Upstream changes from both EDA and hardware developers are regularly incorporated into the suite, and new architectures will be added as they become available and relevant. The result is a benchmark suite that evolves alongside the open-source community and hardware ecosystem, rather than serving as a static snapshot.

\section{Existing Benchmarks}
\label{sec:existing}
Several prior benchmark suites have each played an important role in the advancement of open-source EDA research, and HighTide is designed to complement their contributions rather than replace them. The OpenROAD Flow Scripts (ORFS) benchmark collection~\cite{orfs} pairs a turnkey reference RTL-to-GDS flow and has become the de facto standard for open-source and academic comparisons. The benchmarks within it lean heavily on CPU-centric and legacy architectures. Of the 27 hardware blocks, 16 are RISC-V-related, and even within those sixteen, much of the diversity is superficial. Several of the cores (Ariane, CVA6, BlackParrot v1) are functionally similar single-core RV64GC (RISC-V 64-bit general + compressed) implementations that exercise overlapping flow stress patterns rather than distinct architectural axes. The non-RISC-V designs are concentrated at the small end of the size spectrum (AES, JPEG, GCD, dynamic\_node) and provide little stress to placement (i.e., limited macros) or routing, while the only macro-heavy design in regular use is a single NVDLA partition. 

These same ORFS designs are propagated into important ML infrastructure. For example, CircuitOps~\cite{circuitops} is expressly built to make generative-AI circuit optimization accessible to ML researchers and utilizes designs directly from the ORFS collection, plus one additional NVDLA benchmark. In doing so, it inherits the limited diversity along with the dataset. EDALearn~\cite{edalearn} was the first benchmark to deliver a reproducible, comprehensive RTL-to-signoff flow specifically aimed at democratizing ML-for-EDA research, but has similar limitations. Most designs have cell counts below 100k with only the Berkeley Out-of-Order Machine (BOOM)~\cite{chipyard} exceeding this threshold. More importantly, over a third of EDALearn's designs come from OpenCores~\cite{opencore}, a largely unmaintained database.

These benchmark suites, which remain static, risk becoming stale as designs that were once representative of modern hardware gradually lose relevance. EDALearn has not been updated since its initial release, a trend also seen in designs from ORFS (Table~\ref{tab:OpenROADCompare}). Both suites handle higher-level design languages only superficially. ORFS includes the Chisel-derived tinyRocket (from Rocket Chip~\cite{rocketchip}), and EDALearn ships several BOOM~\cite{chipyard} configurations also written in Chisel, but in both cases only a static generated Verilog snapshot is tracked, so upstream Chisel changes and bug fixes are never re-integrated. The Chisel language itself has also evolved substantially since these snapshots were produced, meaning the static Verilog no longer reflects how the same designs would be elaborated today. Python-based generators are absent from both suites entirely. BlackParrot~\cite{bp} in ORFS further illustrates this stagnation: the most recent update in ORFS is from 2019, while the development team released a new version in 2024 and continues active development to date. These are not trivial changes and added compressed instructions, bit manipulation, and support for multi-core configurations above 16 cores, none of which are reflected in any benchmark suite that uses BlackParrot today. The lack of diversity and recency in these suites will pose a challenge for ML-driven EDA and continued tool development, which requires diverse and contemporary training data for reproducible and generalizable results~\cite{chhabria_ispd25}.

\begingroup
\begin{table}[htbp]
\resizebox{\columnwidth}{!}{
\begin{tabular}{| c | c | c |}
 \hline
 \textbf{Design Name} & \textbf{Last Updated by Authors} & \textbf{Last Tracked by ORFS}\\ [0.5ex]
 \hline
 ariane~\cite{ariane_cva6} & Sep. 2025 & Sep. 2021  \\
 \hline
 black\_parrot~\cite{bp} & Nov. 2025 & Dec. 2019 \\
 \hline
 microwatt~\cite{microwatt} & Dec. 2025 & Apr. 2022 \\
 \hline
 jpeg~\cite{jpeg} & Jun. 2009 & Jul. 2021 \\
 \hline
 cva6~\cite{ariane_cva6} & Feb. 2026 & Jun. 2025 \\
 \hline
 aes~\cite{aes} & Mar. 2009 &  May 2020 \\
 \hline
 dynamic\_node~\cite{dynamicnode} & Jul. 2021 &  Jul. 2021 \\
 \hline
\end{tabular}
}
\caption{OpenROAD design update analysis using a subset of designs reflecting the CPU-centric and legacy-dominated design collection}
\label{tab:OpenROADCompare}
\end{table}
\endgroup


Contest benchmarks have historically driven measurable progress on individual flow stages and serve as the primary form of evaluation for new EDA techniques~\cite{ispd_contests, mlcad_contests}. The ISPD 2024 global-routing contest~\cite{ispd24_paper}, widely used to benchmark large-scale routing innovations, features only one non-RISC-V design, NVDLA~\cite{nvdla}, among its eight benchmarks despite being ``tailored for industrial-level circuits.'' The MLCAD 2025 ReSynthAI contest~\cite{mlcad25}, the first to specifically frame AI-driven logic resynthesis as a learning task follows a similar trend. Four of its seven public benchmarks are NVDLA partitions, with the remainder being smaller legacy designs gathered from the IWLS 2005 benchmarks~\cite{iwls2005}. A contest centered around AI-driven logic resynthesis stands to gain little from training on decades-old designs. It is our hope that HighTide can be used for future contests at both of these conferences.

Beyond design diversity, many contest benchmarks~\cite{ispd_contests, mlcad_contests, iccad_contests} and ML datasets~\cite{circuitnet, circuitnet2, circuitops} are largely ``point-tool'' focused. This means that they typically target individual stages such as clock synthesis, routing, or placement rather than end-to-end evaluation.  CircuitNet~\cite{circuitnet} and CircuitNet 2.0~\cite{circuitnet2} exemplify this as they are built to give ML researchers a training and evaluation set for individual flow stages, and not end-to-end evaluation. They nevertheless derive their data from a narrow set of designs consisting of six RISC-V CPUs at 28nm for CircuitNet, with more advanced blocks like Vortex~\cite{vortex} and NVDLA added on a 14nm process node in CircuitNet 2.0. These point-tool benchmarks, combined with their limited design diversity, do not capture the interactions between flow stages, meaning that improvements demonstrated on a single stage may not translate to end-to-end benefits.

\section{Suite Overview}
\label{sec:overview}
HighTide consists of a diverse set of designs typically seen in commercial systems, listed in Appendix~A. The suite is designed to address each of the gaps identified in Section~\ref{sec:existing} by reducing the RISC-V CPU share from roughly 59\% in ORFS to roughly 30\% in HighTide, with the remaining cores spread across distinct architectural axes (Snitch Cluster's 4-core FPU-equipped compute cluster, BlackParrot v2's multi-core variants, and Minimax's bit-serial micro-coded RV32IC) rather than repeating the same CPU shape, and the freed allocation goes to ML accelerators, NoCs, GPGPUs, peripheral controllers (PCIe, DRAM, Ethernet), and cryptographic datapaths. The ML accelerator category alone spans Gemmini's systolic-array DNN, NVDLA-small's partitioned deep-learning accelerator, two distinct CNN accelerators (NNgen's CIFAR-10 inference datapath and Eyeriss v2's sparse CNN with im2col/GEMM processing elements), Coral NPU's scalar-only inference engine, and Ternip's ternary-precision matrix-multiplication unit. The language coverage spans Verilog, SystemVerilog, Chisel, and two Python-based generators (VerilogGen~\cite{pyverilog}, MiGen~\cite{migen}). The cell-count distribution stretches from under \lowerboundcell to over \upperboundcell so that placement, routing, and timing closure are exercised at multiple scales. Similarly, macro-heavy designs (Eyeriss v2, NVDLA-small, BlackParrot, NyuziProcessor, CNN) are first-class members of the suite rather than outliers. 

Unlike suites that capture a single static snapshot and are never updated, HighTide uses Git submodules to provide stable, versioned snapshots while regularly incorporating upstream changes. For designs that require pre-processing, we also track the versions of supporting tools for replicability. In contrast to point-tool benchmarks, HighTide provides full high-level language (e.g., Chisel or Python) or standard RTL-to-GDS benchmarks that exercise the entire tool flow, including synthesis, floorplanning, placement, clock tree synthesis, routing, and sign-off. 

\subsection{Design Evaluation}
\label{subsec:designeval}
HighTide aims to continuously expand the heterogeneous distribution of designs by adding new ones as they emerge in the open-source community. The initial suite reflects the current landscape yet designs will be incorporated based on the following criteria:
\begin{itemize}
    \item Sufficient documentation, testing, and verification infrastructure.
    \item Signs of continued development and maintenance.
    \item Distinctiveness in comparison to existing suite designs.
    \item Relevance to components found in commercial SoCs.
    \item Adoption within the open-source hardware community.
\end{itemize}
These criteria apply not only during design intake but also on an ongoing basis: a design that goes inactive or loses community relevance may be flagged for removal or replacement in a future release. 

\subsection{Suite Environment}
\label{subsec:environment}
\begin{figure}[htpb]
  \centering
  \includegraphics[width=\columnwidth]{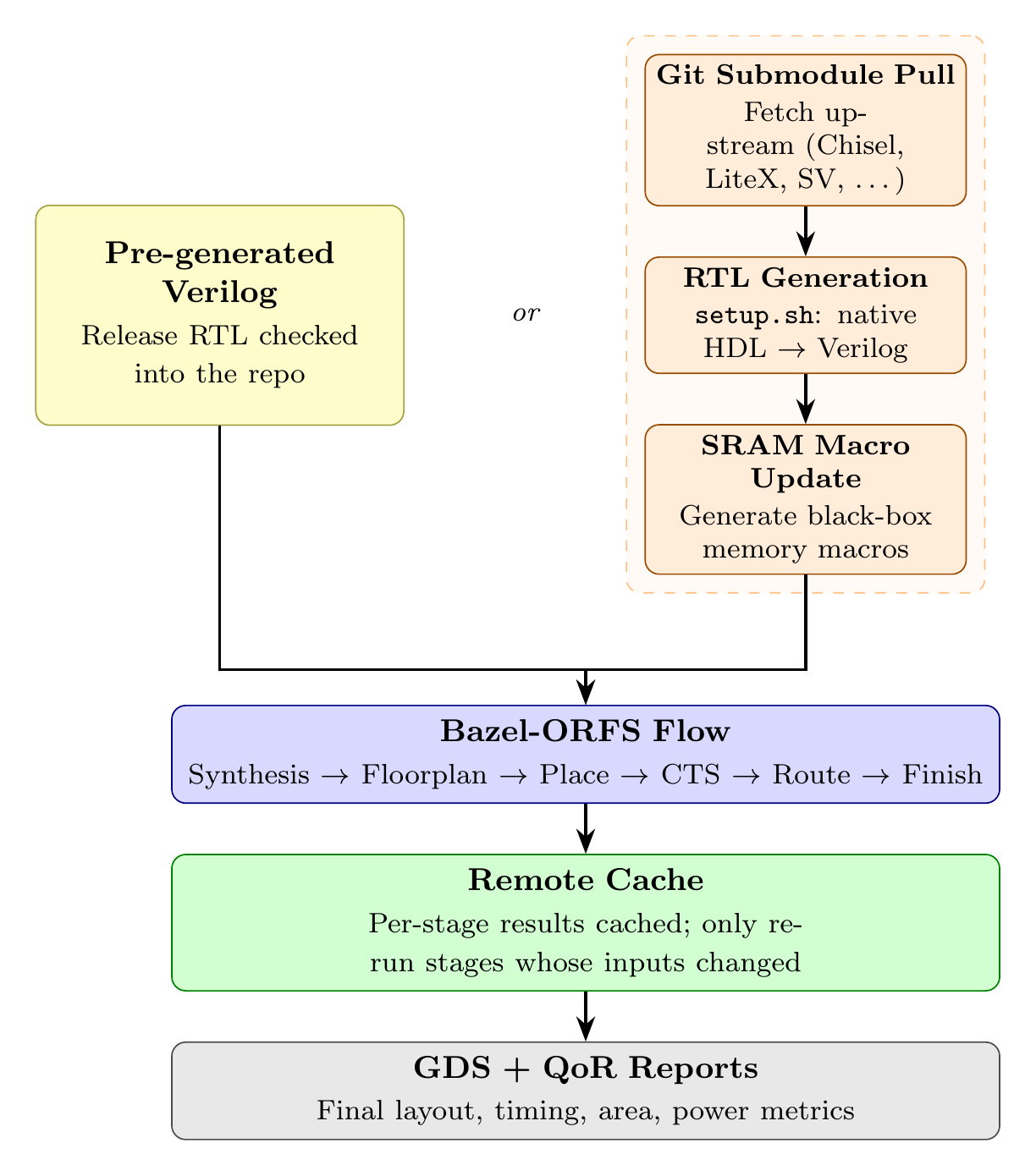}
    \caption{HighTide workflow. The flow takes one of two inputs (pre-generated Verilog committed to the repo, or an optional design-update sub-flow that pulls upstream sources, regenerates Verilog from the design's native HDL, and refreshes black-box memory macros), then runs the unified Bazel-ORFS pipeline (synthesis through finish), backed by a remote per-stage cache, to produce GDS and QoR reports.}
    \label{fig:suite_arch}
\end{figure}
The suite environment distinguishes HighTide, as it ensures tracking as designs evolve and allows integration of new designs without compromising stability. {\bf We plan to make semi-yearly tagged releases so that researchers can do apples-to-apples comparisons.} However, this interval will be adapted to balance meaningful upstream changes with the stability required for reproducible evaluation of a design. We also provide quality-of-results (QoR) tracking of the baseline for easy comparison. To make these releases reproducible, each design ships with a fixed set of configurations, including Synopsys Design Constraints (SDC), memory macro definitions, and power delivery specifications. These configurations will also be updated as tools and designs change.

HighTide is built on top of ORFS~\cite{orfs} via \textit{bazel-orfs}~\cite{bazelorfs}, an existing Bazel integration for ORFS maintained by the OpenROAD project. \textit{bazel-orfs} builds OpenROAD, OpenSTA, Yosys, and the yosys-slang plugin from pinned upstream commits using a hermetic Bazel-managed LLVM toolchain, so users do not need to install the EDA tools separately or run a Docker daemon at execution time. The flow is partitioned into separate build targets that allow re-execution of stages only whose inputs have changed, drastically reducing runtime for large designs. A public remote cache, served by a \textit{bazel-remote} instance, allows users to fetch pre-built baseline results (including the tool binaries themselves) without local computation. Reads are anonymous, while writes require credentials issued to suite maintainers and contributors.

For distributed execution, HighTide provides Kubernetes integration for parallel builds across designs and platforms. Each Kubernetes job runs in a lightweight Ubuntu container, installs Bazel, and invokes \textit{bazel-orfs} exactly as the local flow does. The EDA tools are built from source by Bazel inside the container (or fetched from the remote cache on a hit), with build artifacts shared via the same cache. The configuration is portable to any managed Kubernetes services such as Google Kubernetes Engine (GKE) or Amazon Elastic Kubernetes Service (EKS).

RTL compilation checks are performed on upstream updates, catching integration issues before tagging the collection of designs for release. On failure, suite maintainers are notified that manual intervention is necessary through a manual hotfix, deferring updates for a specific design until the issue is resolved. Each design includes Verilog RTL, pre-processed from its upstream source via a dedicated setup script. This enables support across a range of design languages and supporting hotfixes when necessary. By default, the suite passes this pre-processed Verilog through the RTL-to-GDS flow. A \texttt{bazel build --define update\_rtl=true} build-time flag is also available, which executes the setup script that pulls from upstream sources and recompiles the Verilog to replace the pre-existing RTL (see Figure~\ref{fig:suite_arch}).

The setup scripts are technology-independent, except where technology-specific memory macros limit portability. We use a custom memory generator derived from bsg\_fakeram~\cite{bsg_fakeram} and FakeRAM 2.0~\cite{fakeram2} to replace SRAMs; it provides process design kit (PDK)-specific LEF and Liberty models. Specifically, we modified it to retain the underlying technology constants while adding improved peripheral area estimates as well as multi-port and write-mask support. The generator is not integrated into the \texttt{update\_rtl} command, but it is available as a submodule in the suite so that users can generate memory macros for any supported PDK, without needing to source the tool separately.

Designs in HighTide are supported across three technology platforms with deliberately different physical characteristics. ASAP7~\cite{ASAP7} is a 7~nm predictive academic FinFET node with a large standard-cell library (high-V$_t$, low-V$_t$, and regular-V$_t$ variants and a wide range of drive strengths) and a deep metal stack (M1 through M9 in typical academic flows), making it representative of advanced-node tradeoffs where placement density and timing repair dominate. NanGate45~\cite{nangate45} is a 45~nm planar PDK (the NanGate Open Cell Library on FreePDK45) and behaves as a typical mid-node baseline, with a moderately sized cell library and a balanced metal stack; it is the closest of the three to the conditions that most published EDA experiments still report against. SkyWater 130nm~\cite{sky130hd} is a 130~nm planar foundry-fabricable PDK with substantially fewer usable routing layers than the other two, which makes it especially sensitive to congestion, macro halos, and clock-tree repair on the macro-heavy designs in the suite.  Cross-platform porting requires platform-specific clock period scaling, memory macro regeneration with appropriate metal layer stacks, and calibrated design-rule parameters. Supporting all three of these process points (advanced FinFET, typical planar, and routing-constrained fabricable) reduces the risk that benchmark results are artifacts of a single PDK's characteristics and provides training diversity for ML models.
\begin{figure*}[t]
  \centering
  \begin{subfigure}[b]{0.32\textwidth}
    \centering
    \includegraphics[width=\textwidth]{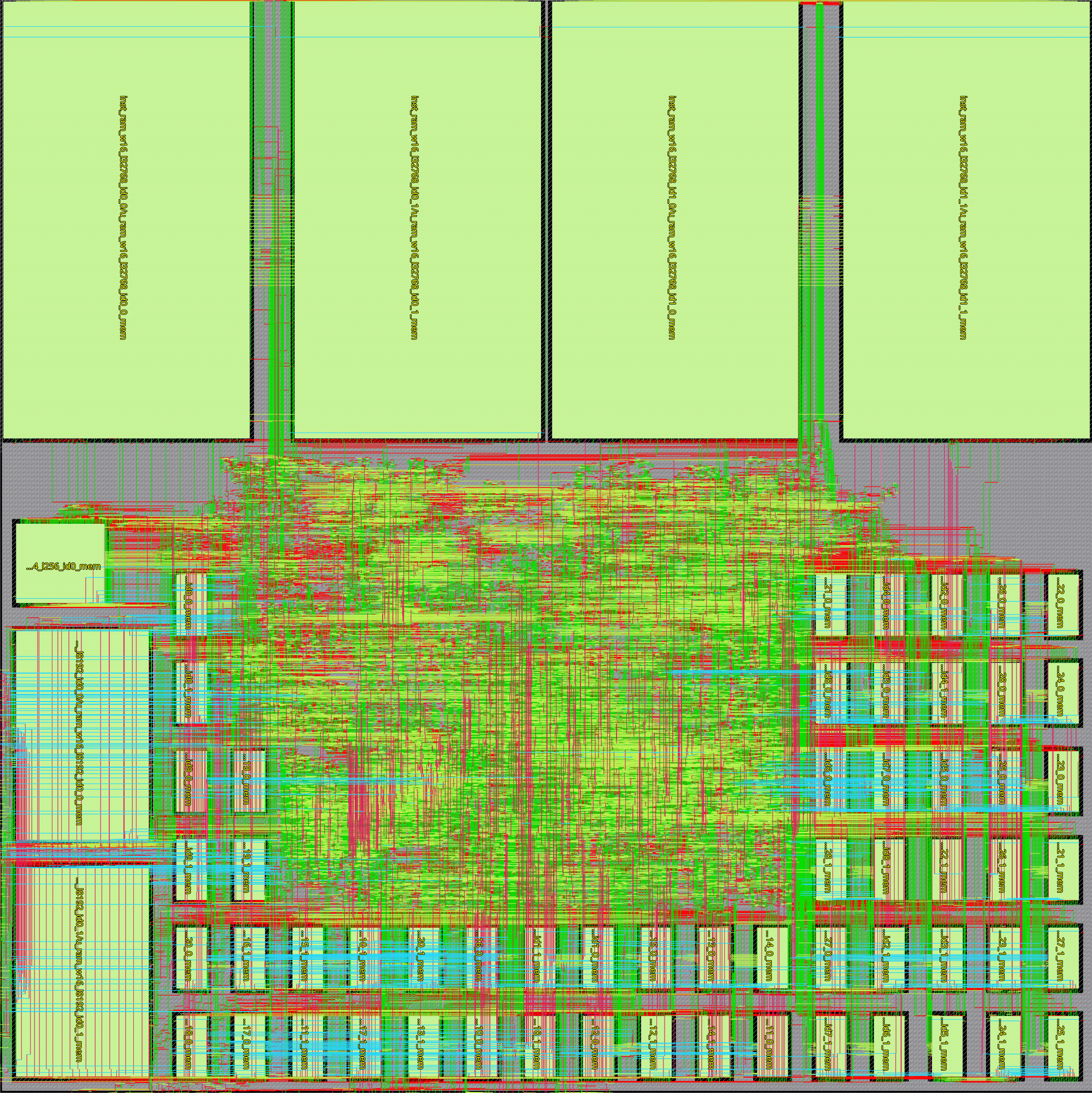}
    \caption{CNN (169k cells, 65 macros)}
    \label{fig:placement_cnn}
  \end{subfigure}
  \hfill
  \begin{subfigure}[b]{0.32\textwidth}
    \centering
    \includegraphics[width=\textwidth]{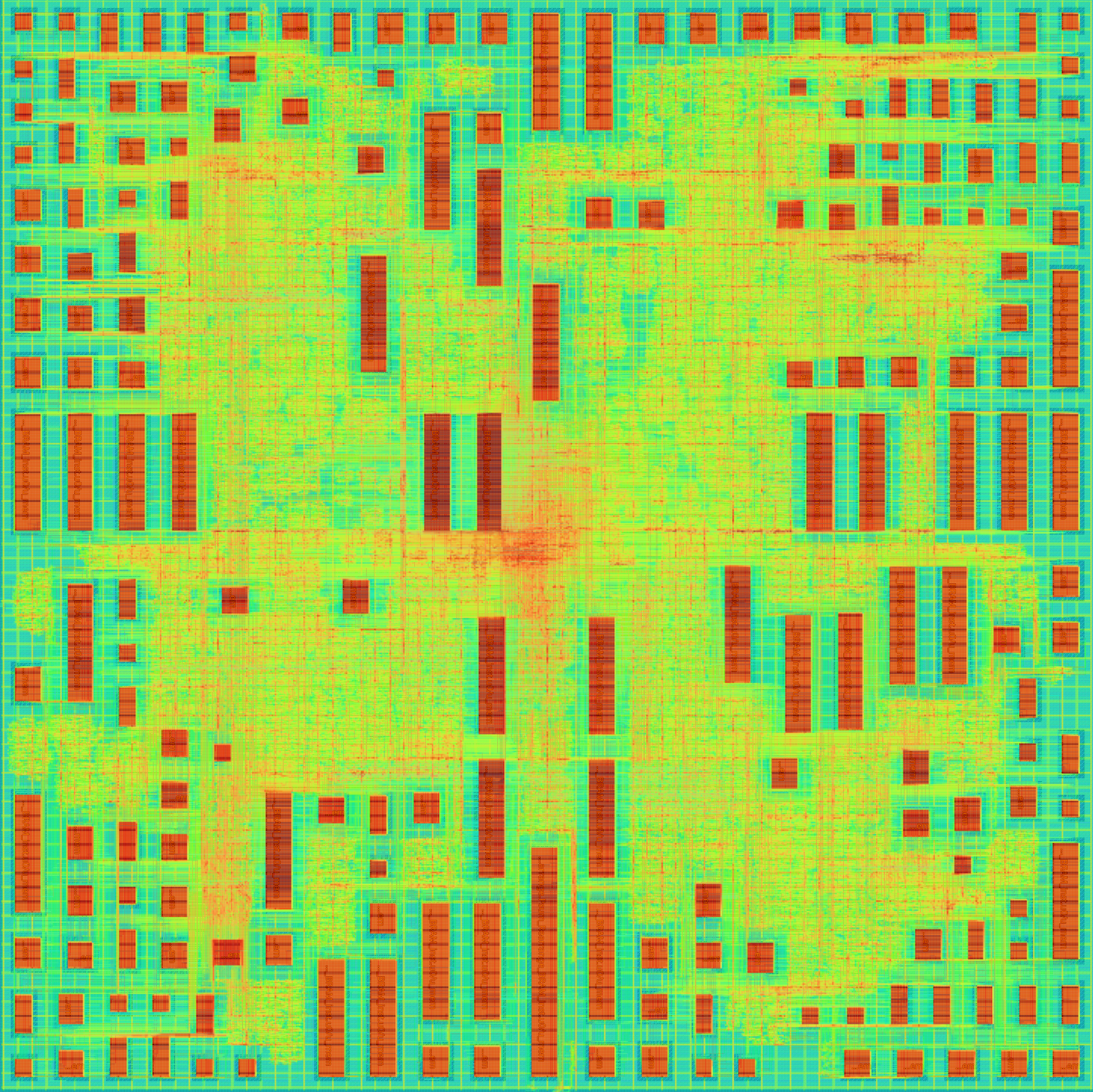}
    \caption{Eyeriss v2 (300k cells, 193 macros)}
    \label{fig:placement_eyeriss}
  \end{subfigure}
  \hfill
  \begin{subfigure}[b]{0.32\textwidth}
    \centering
    \includegraphics[width=\textwidth]{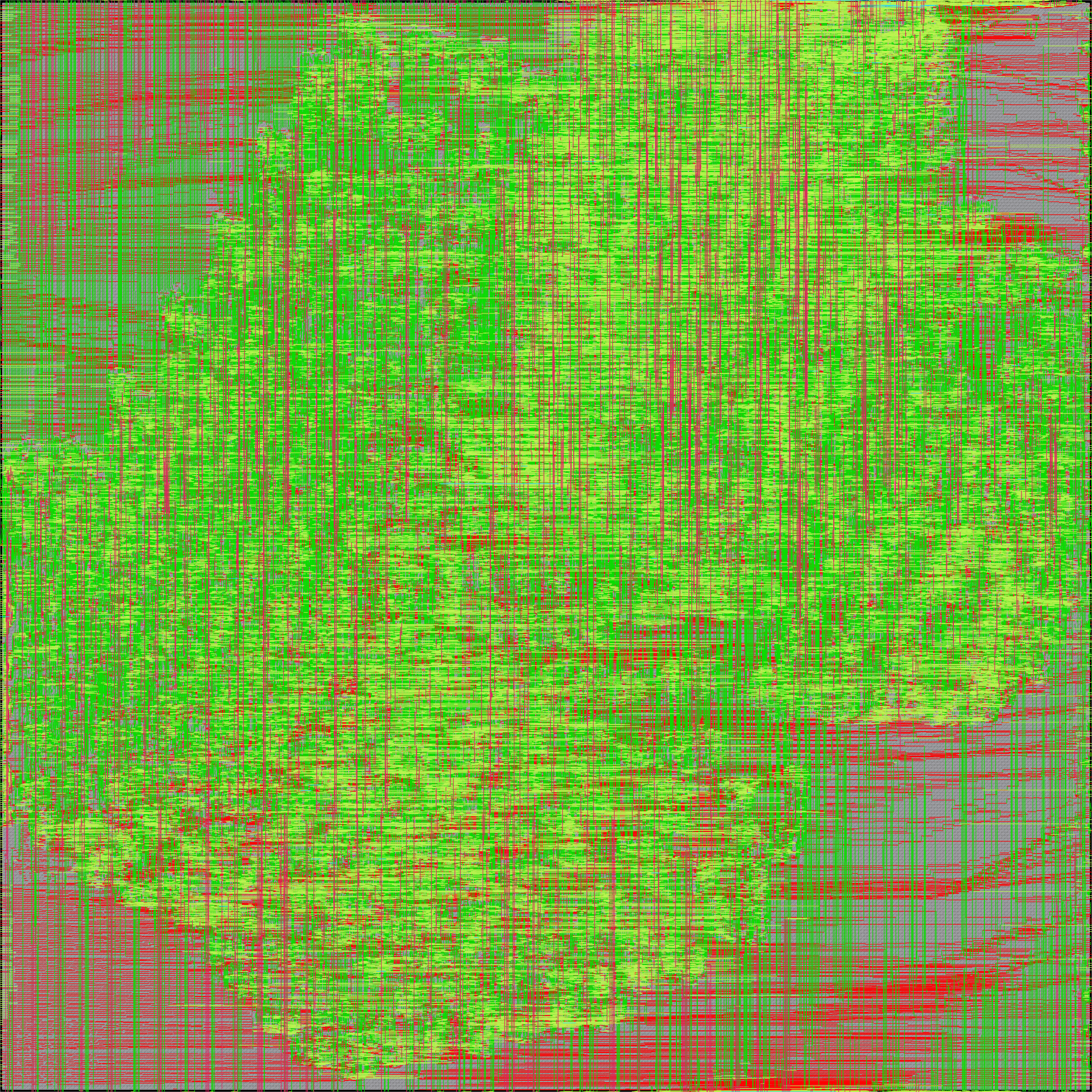}
    \caption{Gemmini (709k cells, no macros)}
    \label{fig:placement_gemmini}
  \end{subfigure}
  \caption{Placement view for three HighTide designs (ASAP7): a small CNN accelerator with dense macros, the most macro-dense design in the suite (Eyeriss v2 with 193 SRAM macros), and the largest macro-free design in the suite (Gemmini).}
  \label{fig:placements}
\end{figure*}

\section{Design Variation}
\label{sec:variation}
The initial release of HighTide covers a wide range of design complexities and applications, with cell counts ranging from under \lowerboundcell to over \upperboundcell as shown in Appendix~A.

Beyond application diversity, designs also vary in their structural composition. Figure~\ref{fig:placements} shows placement views for three designs that highlight this variation. CNN packs 65 large SRAM macros around the NNgen-generated CIFAR-10 inference datapath, producing a layout dominated by regular macro banks that constrain standard-cell placement to the remaining channels. Eyeriss v2, the most macro-dense design in the suite, distributes 193 smaller SRAM macros (input-activation, weight, and partial-sum scratchpads per processing element, plus the global buffer banks) across a 2x2 cluster-group array, exercising both macro placement and routing congestion at a density that none of the other suite designs reach. Gemmini, the largest macro-free design in the suite, tests how the flow scales with large, regular systolic arrays which are somewhat visible in placement clusters.

\begin{figure}[htpb]
  \centering
  \includegraphics[width=\columnwidth]{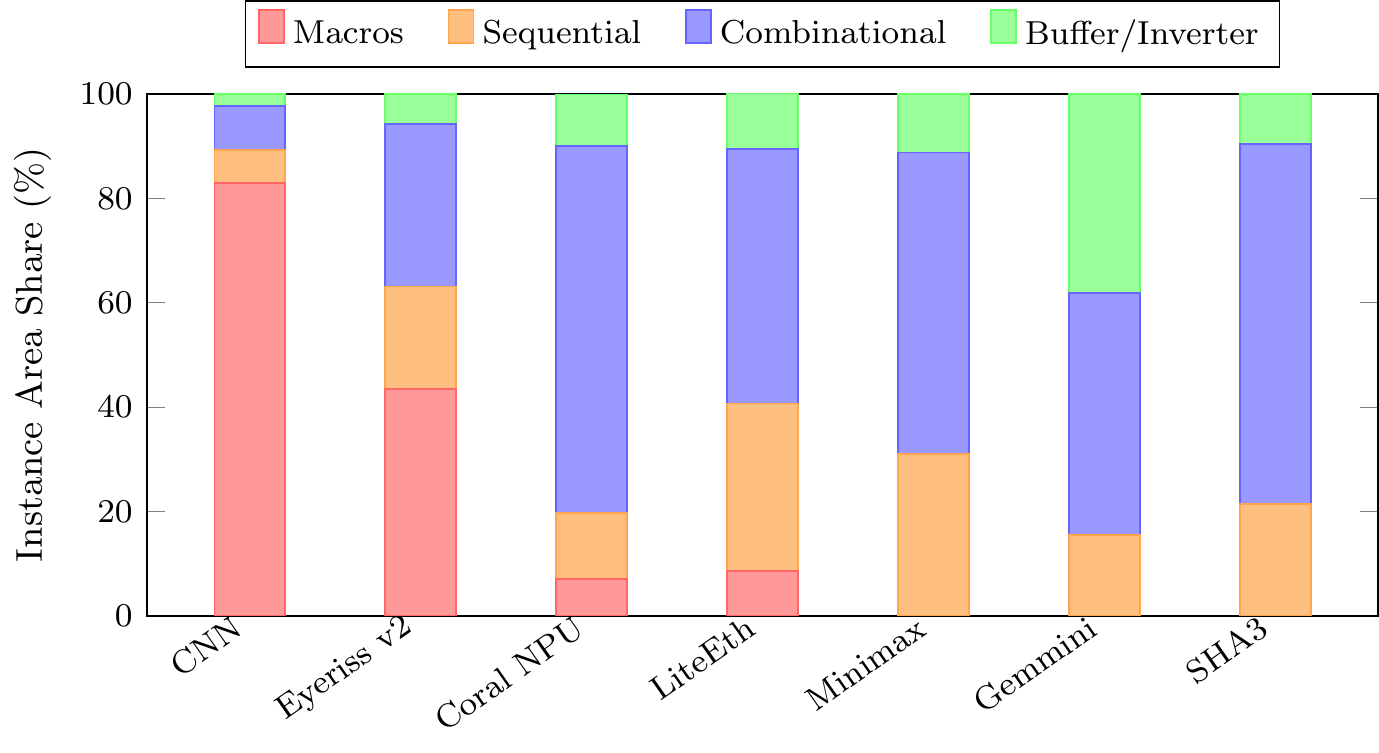}
    \caption{Instance area share (ASAP7) for a selection of HighTide designs. Each bar is normalized to 100\% of the design's reported instance area, partitioned into macro, sequential, combinational, and buffer/inverter contributions. Cell-class areas are estimated as per-class average cell area times count while macro area is the residual against the reported total instance area. }
    \label{fig:areacomp}
\end{figure}

Figure~\ref{fig:areacomp} quantifies these structural differences using an area breakdown across a selection of HighTide designs on ASAP7 by partitioning each design's instance area into macros, sequential cells, combinational cells, and buffer/inverter cells.  Each bar is normalized to 100\% of the design's instance area so that the structural profile is directly comparable across the three orders of magnitude in absolute size between the smallest and largest design (absolute sizes appear in Appendix~A). This single view captures the full range of structural variation including macro dominance (CNN at 83\% and Eyeriss v2 at $\approx$44\% macro area), sequential-heavy designs that stress clock-tree synthesis and hold-timing closure (LiteEth, Minimax), combinational-heavy datapaths that emphasize logic optimization and routing (SHA3, Coral NPU), and the unusually high buffer/inverter fraction in Gemmini ($\approx$38\%) caused by long systolic-array fan-out paths that the resizer must repair. Existing suites dominated by RISC-V CPUs and legacy IP offer limited exposure to this spread of structural profiles.

This variation among designs ensures that flow stages encounter fundamentally different workloads across the suite. As HighTide grows, we plan to expand coverage within categories that we've already represented, such as CPUs with non-RISC-V ISAs, or accelerators targeting different workloads than those currently in the suite.

\section{AI-Assisted Design Curation}
\label{sec:skills}
Maintaining a benchmark suite requires continuous effort in discovering candidate designs, integrating them into the flow, porting across technology platforms, tracking upstream updates, and tuning design configurations. To reduce this burden, HighTide employs Claude Code~\cite{claudecode} skills, which are structured prompts that guide an AI coding agent through multi-step hardware design workflows. Each skill encodes domain-specific knowledge about the suite's conventions, flow requirements, and design configuration patterns, enabling repeatable execution of tasks that would otherwise require significant manual effort and EDA expertise. The suite currently includes twelve skills, organized into four categories: design lifecycle, flow optimization, tool reference, and meta-maintenance. The skills are backed by a shared long-term memory in the form of per-design decision logs (Section~\ref{subsec:decisions}) that allow successful tuning patterns to propagate across designs and platforms.

\subsection{Design Lifecycle Skills}
Four skills manage the end-to-end lifecycle of benchmark designs:

\textbf{Design Discovery.} The \texttt{find-designs} skill searches for open-source hardware designs that meet HighTide's intake criteria (Section~\ref{subsec:designeval}). Given an optional category focus, it evaluates candidate repositories against requirements (open-source license, synthesizable RTL, and verification infrastructure) and opens GitHub issues proposing viable additions. We have already integrated suggestions from this skill and have several candidate additions pending.

\textbf{Design Integration.} The \texttt{new-design} skill incorporates a candidate into HighTide by analyzing the upstream design's HDL language and memory requirements, creating the Git submodule and RTL generation scripts (supporting Chisel/Scala, LiteX/Python, VerilogGen/Python, SystemVerilog via sv2v or yosys-slang, and plain Verilog), generating abstract memory macro LEF/LIB files, creating platform-specific configurations, and validating through the RTL-to-GDS flow.

\textbf{Cross-Platform Porting.} The \texttt{port-design} skill ports designs between technology platforms (ASAP7, NanGate45, SkyWater 130nm). It uses calibrated scaling ratios for initial clock periods and die areas, generates platform-specific memory macros with appropriate metal layer stacks, and adapts power delivery and IO placement configurations.

\textbf{Upstream Tracking.} The \texttt{update-design} skill audits all designs and tool dependencies for upstream changes. It compares each submodule's pinned commit against the upstream version, classifies changes by severity, and handles five update types: source refresh, tool dependency updates, flow parameter tuning, memory macro additions, and cross-platform porting.

\subsection{Flow Optimization Skills}
Three skills assist with design tuning and troubleshooting:

\textbf{Debug Design.} The \texttt{debug-design} skill diagnoses failing builds by analyzing logs and design configurations. It handles synthesis failures (memory inference, missing modules), floorplan and placement issues, routing congestion, timing violations, and DRC errors. The skill can generate layout visualizations including placement density heatmaps and routing congestion maps for diagnosis.

\textbf{Optimize PPA.} The \texttt{optimize-ppa} skill improves power, performance, and area (PPA) metrics. It incrementally increases core utilization while monitoring congestion and DRC violations, tightens clock periods to find the maximum achievable frequency, refactors memory macros by splitting wide FakeRAM banks into narrower ones or merging narrow banks into wider ones when macro geometry (rather than standard-cell density) limits routability, and analyzes IR drop for power optimization. Runtime monitoring serves as an early-warning signal for unviable configurations. This skill and \texttt{debug-design} share a small set of reference skills (timing analysis, congestion analysis, SRAM repartitioning, and layout-image generation) that each invokes as needed, keeping platform-specific heuristics in one place rather than duplicated across skills.

\textbf{Generate SRAM.} The \texttt{generate-sram} skill produces abstract memory macros (LEF, Liberty, and behavioral Verilog views generated by a configurable academic memory model generator) for arbitrary depth/width and port configurations on each supported PDK. For small SRAMs, these are synthesized as DFFs rather than SRAMs. The generator encodes the per-platform technology parameters (metal pitches, manufacturing grid, contacted-poly pitch) and is invoked automatically.

\subsection{Tool Reference Skills}
Two skills provide complete access to EDA tool documentation:

\textbf{OpenROAD Commands.} The \texttt{openroad-tcl} skill provides a cached reference for OpenROAD TCL commands across all modules (global placement, detailed placement, clock tree synthesis, detailed routing, power delivery, etc.), enabling the agent to write correct Tcl scripts without network access. 

\textbf{SDC/STA Commands.} The \texttt{sdc-sta} skill mirrors the OpenSTA user guide, providing command reference for SDC and static timing analysis (STA), used when creating or modifying \texttt{constraint.sdc} files.

\subsection{Meta-Maintenance Skills}
Three skills capture institutional knowledge that accumulates as the suite is exercised:

\textbf{Track Upstream Tool Bugs.} The \texttt{track-bug} skill records reproducible defects in upstream tools (OpenROAD, Yosys, yosys-slang, sv2v, ORFS, bazel-orfs) along with the specific workaround applied in HighTide. It searches the relevant upstream issue tracker for matching reports, captures the affected designs and flow stages, and updates the project-level workaround table so that future builds inherit the mitigation rather than rediscovering it. The skill explicitly excludes design-side configuration issues, which are routed to \texttt{debug-design} or \texttt{optimize-ppa}.

\textbf{Publish Results.} The \texttt{update-results} skill keeps the public results table and gallery synchronized with the latest cached build artifacts. Each row in the results table is pinned to the commit SHA that last touched the design's RTL, BUILD, or SDC; the corresponding gallery image is named with the same SHA so that table rows and images cannot drift. After a build sweep, the skill detects stale rows, refetches the cached results, regenerates layout images, and rewrites the table. This makes per-design QoR history a first-class artifact of the suite.

\textbf{Improve Skills (``Dreaming'').} The \texttt{improve-skills} skill audits recent debugging, optimization, and porting sessions to extract patterns that should be folded back into the skill set itself. It reviews conversation context, recent commits to design configurations and flow infrastructure, and project memory, then updates the relevant skill or proposes a new one. This step plays the same role as the offline ``dreaming'' phase is increasingly used in agentic systems~\cite{dreaming}, in which an agent consolidates problem-solving traces accumulated during active work into reusable abstractions between sessions. By running this consolidation against HighTide's own session history, the suite's automation improves as the design corpus expands rather than treating each new design as an isolated problem.

\subsection{Cross-Design Decision Memory}
\label{subsec:decisions}
The skills above operate on a shared persistent memory in a form of a per-design decision log at \texttt{designs/src/<design>/DECISIONS.md}, with one section per technology and sub-sections for design variants. Each entry records the \emph{why} behind a non-obvious configuration choice (a clock period derived from a measured \texttt{period\_min}, a non-default \texttt{MACRO\_PLACE\_HALO}, a \texttt{PRE\_CTS\_TCL} script that mitigates a specific upstream bug, or a manual macro placement that was needed to clear congestion), rather than the value itself, which lives in \texttt{BUILD.bazel} or \texttt{constraint.sdc}. Cross-links to the canonical upstream-bug index (in \texttt{CLAUDE.md}) ensure that workaround rationale is recorded once and referenced from each affected design.

The flow-optimization and lifecycle skills consult this memory before performing any new work, surfacing entries from designs of similar shape (same platform, comparable macro density, similar gate composition) as candidate solutions; when a new fix is applied, the originating design is credited. This cross-design transfer is what allows a small team to maintain a growing benchmark across three PDKs without re-deriving each platform's idiosyncrasies for every new addition.

\subsection{Impact on Suite Growth}
These skills have directly contributed to HighTide's growth. The \texttt{port-design} skill was used to port the ASAP7 designs to NanGate45, generating platform-specific memory macros and calibrating clock periods from scaling ratios derived from existing cross-platform designs. Subsequent invocations extended SkyWater 130nm coverage to Gemmini, NVDLA, Vortex, CNN, Coral NPU, and SHA3, multiplying total design-platform combinations without proportionally increasing curation effort. The \texttt{track-bug} skill has documented several upstream defects encountered during porting, including OpenROAD MPL-0040 macro placement failures and an ODB-1200 crash that mandates skipping the split-load fanout move on the SkyWater 130nm Gemmini port. The MPL-00400 failure also lead to an issue filed with OpenROAD. Without these skills, and without the decision memory that links them, each operation would require manual analysis, technology-specific memory macro generation, and iterative re-derivation of fixes already discovered on other designs.

\section{Limitations}
\label{sec:limitations}
The introduction of a new suite infrastructure is not without practical tradeoffs. We discuss the scope of these constraints below.
\subsection{Post-Flow Verification \& Silicon Validation}
Although the intake criteria in Section~\ref{subsec:designeval} requires sufficient verification infrastructure, the maturity of verification varies across designs. A standardized post-flow methodology that covers the suite is improbable, but per-design verification flows are achievable; development and in-suite integration for design verification will be deployed in future iterations of HighTide. 

As of its initial release, assessment of the suite relies on design rule checks (DRC), layout versus schematic (LVS), and STA sign-offs from the ORFS flow. We consider this sufficient for the initial release, where our primary objective is showcasing a diversified design-space that passes RTL-to-GDS sign-offs.

\subsection{Memory Support}
Handling for dense, static random-access memory (SRAM) macros and "register files" remains a difficult challenge in open-source flows. We opt for abstract, black-boxed SRAM macros generated via a modified version of a configurable academic memory model generator. This simplifies portability of designs across available PDKs, at the cost of less accurate PPA approximations. Power estimates are particularly affected for memory-dominant designs. Area and performance are also impacted since these abstract macros are only approximations. Despite these trade-offs, abstract memory models remain the best choice given the diverse memory sizes required by the designs across the suite, and the limited availability of open-source memory compilers that support the full range of memory sizes and configurations required across the suite.

\subsection{Flow Dependency}
HighTide uses ORFS as its in-suite RTL-to-GDS flow, meaning that any limitations in the OpenROAD toolchain are inherited by the designs run through it. Such factors can manifest through constraints in global routing, timing closure, and macro handling, among other flow-specific behaviors. Despite this in-suite dependency, designs are not locked into ORFS. Since pre-processed Verilog is checked into each design, users can drive the same RTL through alternative open-source flows such as iEDA~\cite{ieda} or other RTL-to-GDS flows with minimal additional configuration.

\section{Conclusion}
\label{sec:Conclusion}
We presented HighTide, a modern and actively updated open-source VLSI benchmark suite spanning a diverse set of designs across multiple HDLs and three technology platforms, with cell counts from under \lowerboundcell to over \upperboundcell. HighTide combines a Bazel-based RTL-to-GDS flow (from-source tool builds, incremental compilation, remote caching, and Kubernetes-based distributed execution) with twelve Claude Code skills spanning design lifecycle, flow optimization, tool reference, and meta-maintenance, all backed by a shared per-design decision memory so that tuning rationale propagates rather than being rediscovered. HighTide is not a one-off collection of designs but an evaluation suite that grows with the open-source hardware and EDA ecosystems.

\newpage
\bibliographystyle{ACM-Reference-Format}
\bibliography{bibtex_cleaned}

\appendix
\begingroup
\begin{table*}[!t]
\centering
{\Large\bfseries Appendix A}\par\vspace{8pt}
\resizebox{\textwidth}{!}{
\begin{tabular}{| c | p{7cm} | c | c | c | c |}
 \hline
 \textbf{Design} & \centering{\textbf{Overview}} & \textbf{Design Language} & \textbf{Cell Count} & \textbf{Macro Count} & \textbf{Fmax (GHz)}\\ [0.5ex]
 \hline
 Snitch Cluster~\cite{snitch} & RISC-V compute cluster: 4 Snitch cores + 1 DMA core with floating-point unit (FPU), tightly-coupled data memory (TCDM), and I-cache & SystemVerilog & 1,500,000 & 38 & 0.17 \\
\hline
 Gemmini~\cite{gemmini} & Deep neural network (DNN) accelerator (16x16 mesh of 2x2 tiles)& Chisel & 710,000 & 0 & 0.85 \\
 \hline
 NVDLA-small partitions~\cite{nvdla} & Scalable and configurable deep learning accelerator & Verilog & 500,000 & 110 & 0.76--2.70 \\
 \hline
 BlackParrot v2 (uno)~\cite{bp} & Single-core RISC-V processor & SystemVerilog & 470,000 & 140 & 0.25 \\
 \hline
 NyuziProcessor~\cite{nyuzi} & Open-source GPGPU with vector floating-point pipeline & SystemVerilog & 400,000 & 55 & 0.69 \\
 \hline
 Eyeriss v2~\cite{eyeriss} & Sparse CNN accelerator with im2col/GEMM processing elements and multi-mode data flow & Verilog & 300,000 & 193 & 0.43 \\
 \hline
 CNN~\cite{nngen} & Generated convolutional neural network (CNN) accelerator (CIFAR-10 32x32 image classifier) generated by NNgen & Python (VerilogGen~\cite{pyverilog}) & 170,000 & 65 & 1.14 \\
 \hline
 Coral NPU~\cite{coralnpu} & Hardware accelerator (scalar-only) for ML inference & Chisel/SystemVerilog & 157,000 & 2 & 0.33 \\
 \hline
 Vortex GPGPU~\cite{vortex} & General-purpose GPU (GPGPU) processor & SystemVerilog & 156,000 & 26 & 0.96 \\
 \hline
 FlooNoC~\cite{floonoc} & Network-on-Chip (NoC): 4x4 narrow-wide mesh router fabric with AXI4 channels & SystemVerilog & 80,000 & 0 & 1.23 \\
 \hline
 LitePCI~\cite{litepci} & PCIe Gen3 endpoint controller (TLP, DMA, MSI, Wishbone bridge) with black-boxed PCIe PHY & Python (MiGen~\cite{migen}) & 75,000 & 21 & 0.22 \\
 \hline
 Ternip~\cite{ternip} & Ternary-precision matrix-multiplication ML inference accelerator & SystemVerilog & 46,000 & 3 & 0.57 \\
 \hline
 LiteDRAM~\cite{litedram} & DRAM controller with Generic SDR PHY & Python (MiGen~\cite{migen}) & 23,000 & 0 & 0.04 \\
 \hline
 SHA3~\cite{sha3-ucb} & SHA-3 (Keccak) cryptographic hash datapath  & Chisel & 19,000 & 0 & 1.13 \\
 \hline
 LiteEth~\cite{liteeth} & Ethernet MAC + UDP/SGMII with USP GTH PHY & Python (MiGen~\cite{migen}) & 17,000 & 5 & 1.31 \\
 \hline
 Minimax~\cite{minimax} & Bit-serial, micro-coded RV32IC CPU & Verilog & 17,000 & 0 & 1.12 \\
 \hline
\end{tabular}
}
\caption*{HighTide benchmark designs spanning three technology platforms (ASAP7, NanGate45, SkyWater 130nm). Fmax is the achieved maximum frequency on ASAP7 after routing.}
\end{table*}
\endgroup

\end{document}